\documentclass[]{pasj01}
\draft

\usepackage{url}

\begin{document}
\Received{\today}
\Accepted{\today}

\title{Discovery of a transient X-ray source Suzaku~J1305$-$4930 in NGC~4945
%
}

\author{Shuntaro \textsc{ide}\altaffilmark{1}}%

\author{Kiyoshi \textsc{hayashida}\altaffilmark{1,2,3}}
\author{Hirofumi \textsc{noda}\altaffilmark{1,2}}%
\author{Hiroyuki \textsc{kurubi}\altaffilmark{1}}
\author{Tomokage \textsc{yoneyama}\altaffilmark{1}}%
\author{Hironori \textsc{matsumoto}\altaffilmark{1,2}}

\altaffiltext{1}{Department of Earth and Space Science, Graduate School of Science, Osaka University, 1-1 Machikaneyama-cho, Toyonaka, Osaka 560-0043, Japan}
\altaffiltext{2}{Project Research Center for Fundamental Sciences, Graduate School of Science, Osaka University, 1-1 Machikaneyama-cho, Toyonaka, Osaka 560-0043, Japan}
\altaffiltext{3}{Japan Aerospace Exploration Agency, Institute of Space and Astronautical Science, 3-1-1 Yoshinodai, Chuo-ku,Sagamihara,Kanagawa252-5210,Japan}

\email{shunta@ess.sci.osaka-u.ac.jp}


\maketitle

\begin{abstract}
 We report the serendipitous discovery of a transient X-ray source, Suzaku~J1305$-$4930,  $\sim$3~kpc  southwest of the nucleus of the Seyfert 2 galaxy NGC~4945.
Among the seven Suzaku observations of NGC~4945 from 2005 to 2011, Suzaku~J1305$-$4930 was detected four times  in July and August in 2010. The X-ray spectra are better approximated with a multi-color disk model  than a power-law model. At the first detection on 2010 July 4--5, its X-ray luminosity was  $(8.9^{+0.2}_{-0.4}) \times 10^{38}$~erg~s$^{-1}$ and the temperature at the inner-disk radius ($kT_{\rm in}$) was $1.12\pm0.04$~keV.
At the last detection with Suzaku  on 2010 August 4--5, the luminosity decreased to $(2.2^{+0.3}_{-0.8}) \times10^{38}$~erg~s$^{-1}$ and $kT_{\rm in}$ was $0.62\pm0.07$~keV. The source was not detected on 2011 January 29,  about six months after the first detection,
with a luminosity upper limit of $2.4\times10^{38}$~erg~s$^{-1}$. We also find an absorption feature which is similar to that reported in Cyg~X-1.
Assuming the standard disk, we suggest that Suzaku~J1305$-$4930 consists of a black hole with a mass of $\sim$10~$M_{\odot}$. The relation between the disk luminosity and $kT_{\rm in}$ is not reproduced with the standard model of a constant inner radius but is better approximated with a slim-disk model.

\end{abstract}

\section{Introduction}\label{s1}


NGC~4945 is a nearby ($\sim$ 3.72~Mpc; Tully et al. 2013) spiral galaxy with an edge-on configuration (the inclination $i \sim$ 78$^\circ$; Ott et al. 2001).  It hosts a heavily ($N_{\rm H}=4.0\times 10^{24}$~cm$^{-2}$) absorbed Seyfert nucleus, which is one of the apparently brightest active galactic nuclei (AGNi) in the hard X-ray band above 20~keV (Guainazzi et al. 2000; Done et al. 2003). The mass of the central black hole is constrained by H$_2$O megamaser observations to be $M_{\rm BH} \sim 1.4 \times 10^6$ $M_{\odot}$ (Greenhill et al. 1997).

Several bright X-ray sources are reported in  NGC~4945 (Kaaret \& Alonso-Herroro 2008; Guainazzi et al. 2000; Schurch et al. 2002), some of which are classified as ultra-luminous X-ray sources (ULXs, as defined in Feng \& Kaaret 2005). A transient ULX was discovered in a Suzaku observation of NGC~4945 in 2006 (Isobe et al. 2008). Its bolometric luminosity ($L_{\rm bol}$) was $4.4 \times 10^{39} \alpha$ erg s$^{-1}$, where $\alpha$ is  ($\cos60^{\circ}$ / $\cos i$), corresponding to the Eddington luminosity for a source of a mass of 20 $M_{\odot}$.
The source  showed the relation of the Stefan Boltzmann's law L$_{\rm bol} \propto T_{\rm in}^4$ (where $T_{\rm in}$ is the temperature at the inner disk radius), which is a typical characteristic of  the standard accretion disk. The innermost disk radius in this scenario is estimated to be $R_{\rm in} =$  $68\pm4$~km,  implying the mass of the black hole to be $M_{\rm BH} =$ 9~$M_{\odot}$, providing that it is a non-rotating  black hole. Isobe et al. 2008 alternatively interpreted this source with emission from a spinning black hole with a 20--130~$M_{\odot}$.

ULXs in nearby galaxies have been  targets of extensive studies.
 Whereas intensive observation  campaigns  for individual sources  have been  performed in our galaxy, the LMC, and SMC,
 intensive observations to such an extent have been  a rarity  for more distant galaxies, except  for several
ULXs, e.g., NGC1313~X-2 (Weng et al. 2014) and Ho~IX~X-1 (Jithesh et al. 2017).
In this paper, we report our serendipitous discovery and frequent monitoring observations of a transient X-ray source in NGC~4945 with Suzaku. It  was nearly ultra-luminous with the  highest observed luminosity of 8.9$\times 10^{38}~$erg~s$^{-1}$.

In this paper, the uncertainties quoted in the tables represent 90\% confidence level  whereas those in the figures and text do 68\%.  Spectral model fitting is performed with the software \texttt{XSPEC} released as a part of the \texttt{HEASOFT} package(version 6.25).
We define $L_{\rm x}$ as the unabsorbed luminosity in the 0.3--10~keV  band and $L_{\rm disk}$ as the bolometric luminosity of the disk.

\section{Observations and Data Reduction}\label{s2}

\renewcommand{\arraystretch}{1}
\begin{table*}[t]
  \caption{List of the Suzaku observations of NGC~4945.}

 \label{all_tbl}
 \begin{center}
  \begin{tabular}{ccccc}
   \hline\hline
    OBSID & Observation date  & Exposure (ksec) &  ID \\\hline
  100008010 & 2005-08-22 & 14.02 & 2005 1st \\
  100008030 & 2006-01-15 & 95.07 & 2006 1st \\
  705047010 & 2010-07-04 & 39.07 & 2010 1st \\
  705047020 & 2010-07-09 & 44.15 & 2010 2nd \\
  705047030 & 2010-07-26 & 40.33 & 2010 3rd \\
  705047040 & 2010-08-30 & 39.39 & 2010 4th \\
  705047050 & 2011-01-29 & 46.11 & 2011 1st \\
\hline\hline
  \end{tabular}
\end{center}
\label{tab:observed}
\end{table*}

NGC~4945 was observed seven times with the fifth Japanese X-ray satellite Suzaku; once in 2005, once in 2006, four times in 2010, and once in 2011, as  summarized in table \ref{tab:observed}. We employed the X-ray Imaging Spectrometer (XIS: Koyama et al. 2007) onboard Suzaku, consisting of three Front-Illuminated (FI) X-ray CCD cameras  called XIS0, 2, 3 and a Back-Illuminated~(BI) X-ray camera called XIS1. They were operated in the normal full-window mode in all the observations. In this paper, we analyzed the cleaned event datasets (Processing script version 3.0.22.43) of XIS0 and XIS3 prepared  with the standard screenings, which exclude events with an Earth elevation angle  lower than $5^{\circ}$ or Earth day-time elevation angles  lower than $20^{\circ}$, and those in the South Atlantic Anomaly. Although the Hard X-ray Detectors (HXD; Takahashi et al. 2007; Kokubun et al. 2007) were also operated, we did not use their data in the results reported in this paper.

\begin{figure*}[h]
\begin{center}
\FigureFile(110mm,110mm){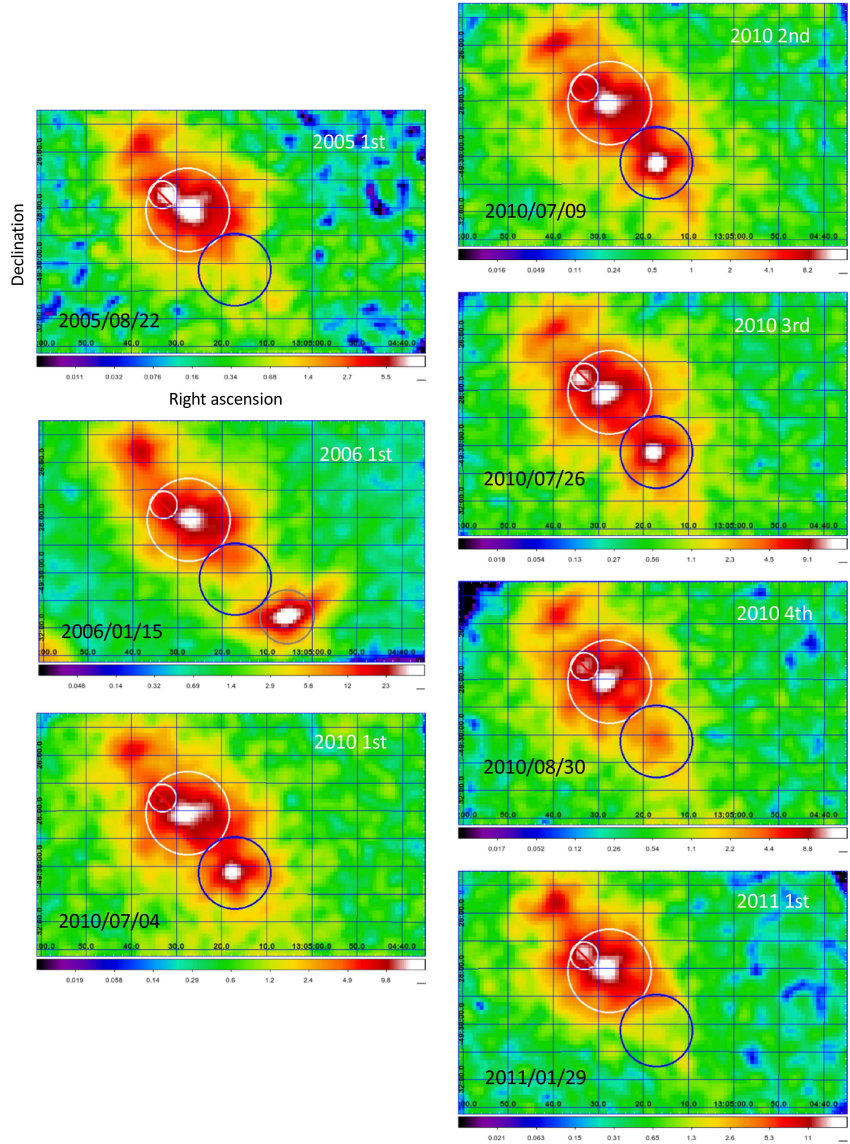}
\end{center}
\vspace{2cm}
\caption{Suzaku XIS0 images of NGC~4945 in the seven observations (table~\ref{all_tbl}). The energy range is 0.2--12~keV. The  gray circle in the 2006 1st image  indicates the position of Suzaku J1305$-$4931 (Isobe et al. 2008). The other circles show the regions from which the spectra are extracted; the AGN-region spectrum in each observation is extracted from the white circular region excluding the smaller white circle in each panel and the SOURCE-region spectrum is from the blue circle. }
\label{fig:all_image}
\end{figure*}

Figure \ref{fig:all_image} shows X-ray images of NGC~4945 obtained by XIS0 in the seven observations. Note that the sky coordinates in the original Suzaku data have offsets by about 0.5$^{'}$. We corrected those offsets by referring {\it Chandra} X-ray images of NGC~4945. In all of them, the active galactic nucleus (AGN) at ($\alpha, \delta$)(J2000) = ($13^{h}05^{m}27^{s}.5, -49^{\circ}28^{'}06^{''}$)  was consistently bright and unresolved emission extending up to about 10~kpc from the central AGN  was detected.
In addition to them, a bright X-ray point source appeared at ($\alpha, \delta$)(J2000) = ($13^{h}05^{m}05^{s}.5$, $-49^{\circ}31^{'}39^{''}$) (marked with a  gray circle in figure \ref{fig:all_image}), about 5 arcmin south-west  of the nucleus, on 2006 January 15--17th, which is the ultra-luminous X-ray source already reported by Isobe et al. (2008).
Moreover, two X-ray sources are found  to the north-east of the nucleus  at ($\alpha, \delta$)(J2000) = ($13^{h}05^{m}32^{s}.9, -49^{\circ}27^{'}34^{''}$) and  to the further north-east  at ($\alpha, \delta$)(J2000) = ($13^{h}05^{m}38^{s}.9, -49^{\circ}25^{'}45^{''}$), both of which were also reported as point sources (Kaaret et al. 2008).
 In all the four observations in 2010, we find an  another X-ray source at  ($\alpha, \delta$)(J2000) = ($13^{h}05^{m}17^{s}.0, -49^{\circ}30^{'}15^{''}$), to the south-west  of the nucleus. As the source is not detected in the other observations,  it is an X-ray transient by definition. To our knowledge, this transient has  never been reported, including the paper on the X-ray spectral change of the NGC~4945 nucleus by Marinucci et al.(2012), and thus it is our discovery. Hereafter, we name the new X-ray transient as Suzaku~J1305$-$4930. The Chandra and XMM-Newton satellite observed NGC~4945 seven and two times, respectively. However, Suzaku~J1305$-$4930  was not detected in any of them.

 The source events  of Suzaku~J1305$-$4930 are accumulated from the 1$'$.3-radius circle region centered at ($\alpha, \delta$)(J2000) = ($13^{h}05^{m}17^{s}.0, -49^{\circ}30^{'}15^{''}$), which is hereafter  referred to as the ``SOURCE'' region (blue circle in Figure \ref{fig:all_image}).
This ``SOURCE'' region inevitably suffers from  contamination from the bright AGN core.  We define another region named  ``AGN'' region centered at the position of the AGN, the events from which are used to estimate the contaminated component.   It is the 1$'$.5-radius circle centered at ($\alpha, \delta$)(J2000) = ($13^{h}05^{m}27^{s}.5, -49^{\circ}28^{'}06^{''}$) excluding the 0$'$.5-radius circle centered at ($\alpha, \delta$)(J2000) = ($13^{h}05^{m}33^{s}.0,-49^{\circ}27^{'}34^{''})$ to the north-east  of the AGN, where another fainter source is visible in some observations (figure \ref{fig:all_image}).
\newline \indent The Non X-ray Background~(NXB) in the SOURCE and AGN regions are estimated by using  \verb|xisnxbgen| (Tawa et al. 2008) in  HEASOFT in \verb|heasoft|-6.25\footnote{\tt See https://heasarc.gsfc.nasa.gov/lheasoft/} and are subtracted from the source spectra. The Cosmic X-ray Background (CXB)  is taken into account as an \verb|XSPEC| model of a power law modified by an exponential cut-off (\verb|powerlaw*highecut| in \verb|XSPEC|; Boldt 1987) and is incorporated in the models employed in the subsequent analyses. We sum up events from XIS0, 3 (and XIS2 if present) and use them as the FI events. We generated redistribution matrix files (rmf) and auxiliary response files (arf), using \verb|xisrmfgen| and \verb|xissimarfgen| (Ishisaki et al. 2007), respectively.
\newline \indent For  each spectrum extracted from the SOURCE region, we use three types of arf files: one  in which a point source at the position of Suzaku~J1305$-$4930 is assumed, another in which  a point source at the AGN  position is assumed, and the other in which uniform  emission is assumed to emulate the CXB. As for the AGN region, we ignore the contamination from the SOURCE region, given that the apparent brightness of the AGN region is much fainter than that of the SOURCE region, and use two types of arf files, where the assumptions are the same as in the latter two of the above-mentioned three arf files. We thus use five arf files in total.

\section{Data Analysis and Results}\label{s4}

\subsection{AGN-region spectra}\label{subsec:agn}
\begin{figure*}[h]
\begin{center}
\FigureFile(100mm,100mm){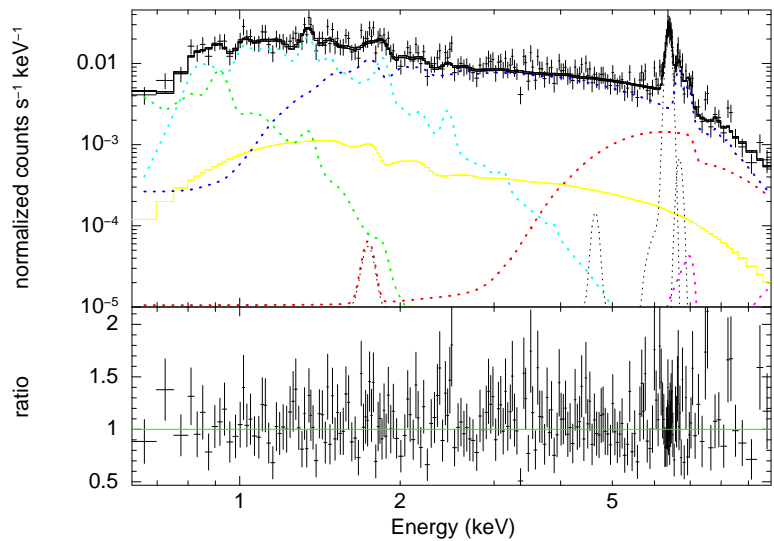}
\end{center}
\caption{The 2011 1st AGN-region spectrum. Green,  cyan and blue dotted lines represent thermal plasma model (\texttt{apec}) with different temperatures. Red line  does a \texttt{pexrav} model. Magenta  does a heavily absorbed power-law model. Yellow  does the CXB model (Section~\ref{s2}). Black dotted lines do three \texttt{Gaussian} lines.}
\label{fig:agn}
\end{figure*}

The emission from the central AGN may significantly contaminate the SOURCE region, given its brightness and proximity compared with the size of
 the point-spread function of the X-Ray Telescopes (XRTs; Serlemitsos et al. 2007). To evaluate the degree and effect of the contamination, we first examine the spectrum extracted from the AGN region in 2011,  the 2011 AGN spectrum, where the new transient source was not detected and hence the contamination of its emission to the AGN spectrum is negligible. The spectrum is binned into energy intervals so that each bin contains 20 counts  at minimum, as shown in Figure \ref{fig:agn}.  It clearly shows a 0.5--10~keV broadband continuum, Fe-K emission lines  in 6.4--7~keV, and multiple emission lines in the 0.5--2~keV band. The soft X-ray continuum of NGC~4945 is most likely to originate in hot plasma located at  outer regions of the AGN and the host galaxy (e.g, Schurch et al. 2002). We  fit the  spectrum with  model based on one of the models employed in Schurch et al. (2002), which consists of a heavily absorbed intrinsic power-law continuum, Compton reflection continuum, three soft thermal continua, and three fluorescence lines  of 6.40~keV Fe K$\alpha$, 7.06~keV Fe K$\beta$, and 7.47~keV Ni K$\alpha$. We replace the  components \verb|mekal| and \verb|wabs| used in Schurch et al. (2002) with \verb|apec| and \verb|tbabs| in our modeling. We also include the galactic absorption  component \verb|tbabs| with the column density fixed at $N_{\rm H} = 1.38 \times 10^{21}$~cm$^{-2}$, the value of which is estimated  with the \verb|nh| command in \verb|heasoft|. In summary, the model  to fit the 2011 1st AGN spectrum is  described in the terminology of \texttt{XSPEC} as \verb|tbabs*(tbabs*powerlaw + tbabs*pexrav + tbabs*apec*3 + zgauss*3)| plus the CXB model described in the previous section.

\begin{figure}[h]
\begin{center}
\FigureFile(120mm,120mm){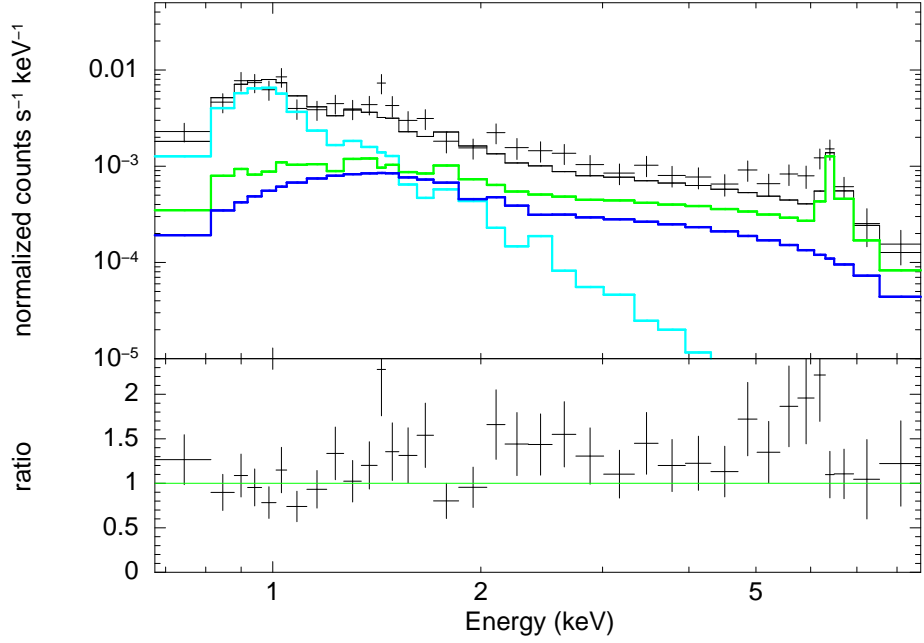}
\end{center}
\caption{The 2010 1st SOURCE-region spectrum. Green: Contamination from the AGN region. Blue: CXB. Cyan: Additional thermal plasma (\texttt{apec}) model intrinsic to th emission from this region.}
\label{fig:source+agn}
\end{figure}

 In the model fitting, the photon index $\Gamma$ and absorbed column for the heavily-absorbed intrinsic power-law component are fixed at 1.55 and 4.0 $\times 10^{24}$~cm$^{-2}$ (Guainazzi et al. 2000), respectively. For the \verb|pexrav| component, the metal abundance is fixed to be the solar value, and the cut-off energy is fixed at 200~keV. The inclination of the reflector is set to 60$^{\circ}$ and the reflection scaling factor is fixed at 0.003 (Itoh et al. 2008).
\newline \indent  This model is found to  reproduce the AGN spectrum successfully.  Table \ref{tab:agn_parameter} lists the best-fit parameters.

\renewcommand{\arraystretch}{1}
\begin{table*}[t]
\caption{Best-fit  parameters  for the AGN-region spectra.}
 \begin{center}
  \begin{tabular}{ccccc}
   \hline\hline
  parameters & 2010 1st & 2010 2nd & 2010 3rd & 2010 4th \\ \hline
  $N_{\rm H,power-law}$~(10$^{22}$ cm$^{-2})$ &	$13^{+6}_{-4}$ &	$9 \pm 3$ &	$7.8^{+0.2}_{-0.6}$ &	$18.2^{+0.4}_{-0.6}$ \\
  $N_{\rm H,1}$~(10$^{22}$ cm$^{-2})^{\rm a}$ &	$<$~0.2 &	$0.4^{+0.3}_{-0.3}$ & $<$~0.5 & $0.5^{+0.2}_{-0.3}$ \\
  $k T_{1}$~(keV$)^{\rm a}$ &	$0.34^{+0.13}_{-0.04}$ &	$0.22^{+0.06}_{-0.04}$ &	$0.3 \pm 0.1$ &	$0.35^{+0.09}_{-0.19}$ \\
  $N_{\rm H,2}$~(10$^{22}$ cm$^{-2})^{\rm b}$ &	$0.9^{+0.2}_{-0.2}$ &	$1.7^{+2.0}_{-0.7}$ &	$1.1^{+1.6}_{-0.9}$ &	$<$~15.7 \\
  $k T_{2}$~(keV$)^{\rm b}$ &	$0.8^{+0.1}_{-0.1}$ &	$0.7^{+0.3}_{-0.3}$ &	$0.9^{+0.2}_{-0.2}$ &	$1.1^{+1.7}_{-0.7}$ \\
  $N_{\rm H,3}$~(10$^{22}$ cm$^{-2})^{\rm c}$ &	$0.9^{+0.4}_{-0.3}$ &	$0.3^{+0.4}_{-0.2}$ & $<$~8.4 &	$1.3^{+2.0}_{-0.9}$ \\
  $k T_{3}$~(keV$)^{\rm c}$ &	$4.1^{+1.3}_{-1.0}$ &	$5.4^{+1.0}_{-1.7}$ &	$6.9^{+4.7}_{-0.5}$ &	$5.2^{+0.2}_{-0.2}$ \\ \hline
  ${\chi}^2$ / d.o.f. & 192.4 / 213 & 201.7 / 185 & 217.6 / 191 & 173.9 / 175 \\
\hline\hline
  \end{tabular}
\end{center}
{\small $^{\rm a}$,  $^{\rm b}$,  $^{\rm c}$:  \texttt{apec} parameters as represented as the components in green, cyan, and blue, respectively, in figure \ref{fig:agn}.}

\label{tab:agn_parameter}
\end{table*}

\subsection{SOURCE-region spectra}\label{subsec:source}

We  analyze the 2011 1st SOURCE spectrum (figure \ref{fig:source+agn}),  for which Suzaku~J1305$-$4930 is not detected significantly.   If  no intrinsic emission  exists in the SOURCE region,  the 2011 SOURCE spectrum  should be reproduced only with the contamination from the AGN and the CXB. We apply a model to the spectrum to examine the hypothesis, assuming the same emission model for the AGN that reproduces the 2011 AGN spectrum summarized in table \ref{tab:agn_parameter}.

We find that a hard X-ray  emission above $\sim$ 4~keV is primarily explained by the contamination from the AGN, while there remains an excess in the soft X-ray band at $\sim$ 1~keV, resulting in an unacceptably large ${\chi}^2$ / d.o.f. = 54.96 / 36. Then, we add a component of thermal emission (\verb|apec| model in \verb|XSPEC|) with a metal abundance fixed at the solar value.  As a result, the fit is improved with ${\chi}^2$ / d.o.f. = 42.8 / 34. However, the fit result is not yet satisfactory and the spectrum above $\sim$ 4~keV shows a slight systematic excess to the model. We  suspect that this potentially originates from the imperfect arf models due to calibration uncertainty and/or attitude  parameters. We thus introduce a multiplicative constant  factor as a free parameter to express the effect, while  all the other model parameters are fixed  to the best-fit parameters already obtained, and refit the spectrum.  Consequently, the fit is further significantly improved with ${\chi}^2$ / d.o.f. = 31.49 / 33. Figure \ref{fig:source+agn} shows the resultant best-fit model overlaid on the data and residuals. The electron temperature of the additional hot plasma and the multiplicative constant are  determined to be 0.88$\pm0.07$~keV and $1.47\pm0.11$, respectively. Hereafter, we employ this model to reproduce the baseline of the SOURCE spectra (in any epoch).

\begin{figure}[h]
  \begin{center}
    \FigureFile(100mm,100mm){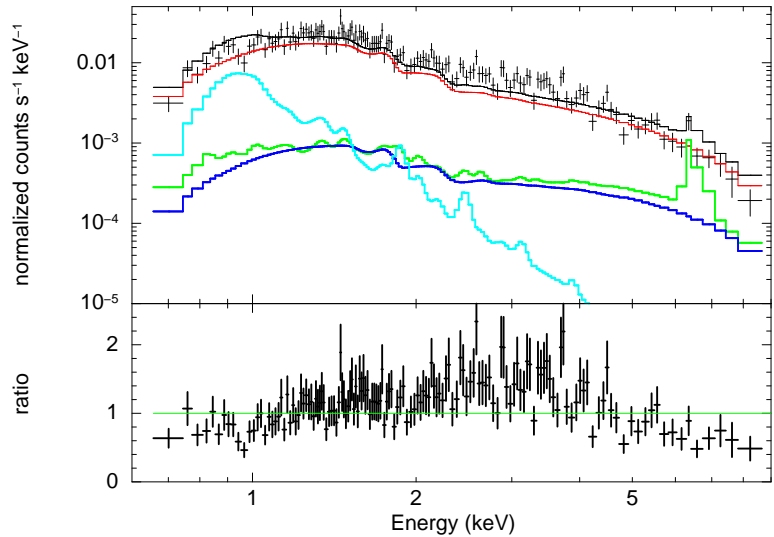}
  \end{center}

  \caption{Power-law model fit to the 2010 1st SOURCE-region spectrum. Red solid line represents the power-law model for the transient source. Cyan, green, and blue  solid lines represent the thermal plasma component intrinsic to the SOURCE-region spectrum, contamination from the AGN region, and the CXB, respectively.}
  \label{fig:7010_pl}
\end{figure}

We then analyze the SOURCE spectra in 2010, in which Suzaku~J1305$-$4930 was significantly detected. Figure  \ref{fig:7010_pl} shows the NXB-subtracted 2010 1st SOURCE spectrum.   An excess emission  is clearly identified in the whole 0.5--10~keV band. The emission is apparently dominated by  that from Suzaku~J1305$-$4930. \newline \indent We first  apply a power-law model modified by the galactic absorption \verb|tbabs| with $N_{\rm H}$ = 1.38 $\times 10^{21}$~cm$^{-2}$ to the spectra.  Figure \ref{fig:7010_pl} shows the fitting result with the 2010 1st SOURCE spectrum as an example.
The fits  are found not to be acceptable; for example, it gives  ${\chi}^2$ / d.o.f. = 313.9 / 145  with the best-fit photon index of $\sim$ 2.04 for the 2010 1st SOURCE spectrum. The residuals show that the spectrum prefers a more concave continuum model than that of a power law.

\begin{figure*}[h]
\begin{center}
\FigureFile(140mm,140mm){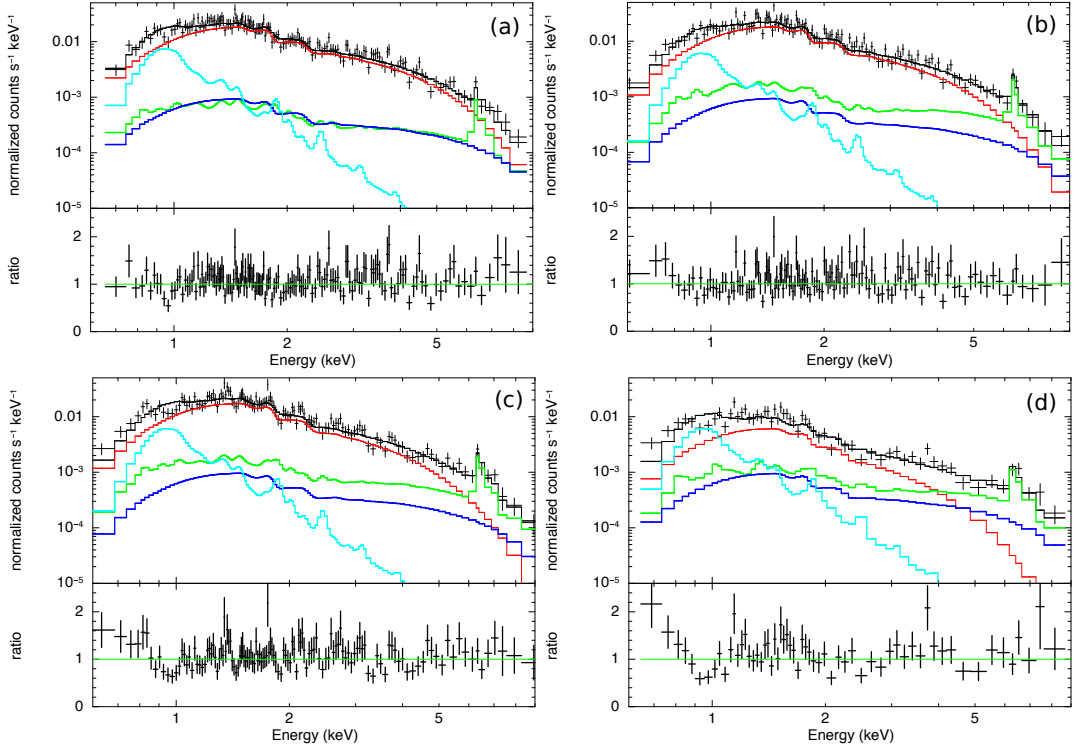}
\end{center}
\caption{MCD model fits to the SOURCE region spectra: (panel a) 2010 1st, (panel b) 2010 2nd, (panel c) 2010 3rd, and (panel d) 2010 4th. Red solid lines represent the MCD model for the transient. Cyan, green, and blue solid lines are the same as those in figure 4. }
\label{fig:dbb_all}
\end{figure*}

We then  apply a modified model, replacing the power law by the multi-color disk black-body (MCD) model (\verb|diskbb| in \verb|XPSEC|; Mitsuda et al. 1984) to the spectra in the epochs where the source was significantly detected, i.e., 2010 1st to 4th.  We find that the MCD model gives acceptable  fits (${\chi}^2$ / d.o.f. = 147.6 / 148 for the 2010 1st SOURCE, for example), unlike  the above-mentioned power-law model. The best-fit parameters for the 2010 1st SOURCE, for example,  are $kT_{\rm in}$ = $1.12\pm0.04$~keV and $R_{\rm in}$ = $64^{+10}_{-9} \alpha$~km for $\alpha^{1/2}$ = $\cos60^{\circ}$ / $\cos i$, where $i$ is the inclination of the accretion disk from  the line of sight.
Figure \ref{fig:dbb_all}  and table \ref{tab:parameter_dbb} show the result of the fitting.

\begin{figure*}[h]
\begin{center}
\FigureFile(140mm, 140mm){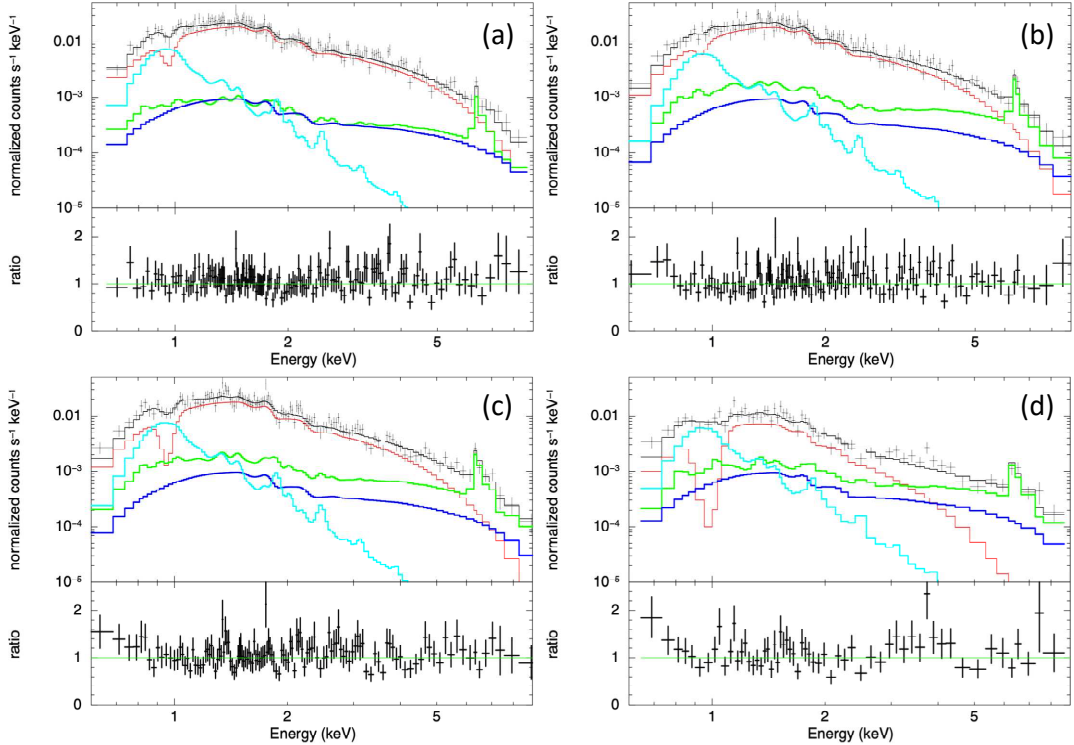}
\end{center}
\caption{SOURCE-region spectra fitted with  the \texttt{gabs*diskbb} model: (panel a) 2010 1st, (panel b) 2010 2nd, (panel c) 2010 3rd, and (panel d) 2010 4th. Coloring of the model components is the same as in figure \ref{fig:dbb_all},  except for the fact that the models here include a Gaussian absorption model. }
\label{fig:gabs_all}
\end{figure*}

Systematic residuals, positive below 0.9~keV and negative around 1~keV, still remain in the spectra (figure \ref{fig:dbb_all}), in particular in the 2010 3rd and 2010 4th spectra. We test two separate additional models: a multiplicative absorption edge model (\verb|edge| in \verb|XSPEC|) and an absorption Gaussian-line model (\verb|gabs| in \verb|XSPEC|), expecting that either or both of them may be more appropriate. Consequently, the fit is significantly improved with the \verb|gabs| model as  summarized in Figure \ref{fig:gabs_all} and table \ref{tab:parameter_gabsdbb}, but not much with the edge model.
\newline \indent Various absorption lines of highly ionized ions of  Ne, Mg, Ne, Fe are observed in the energy band around 1~keV in the X-ray spectra of Cyg X-1 observed with the {\it Chandra} High Energy Transmission Grating (HETG) (Hanke et al. 2009, Nowak et al. 2011, Mi$\check{\rm s}$kovi$\check{\rm c}$ov$\acute{\rm a}$ et al. 2016). The absorption lines are different in different orbital phase and interpreted as absorption in the stellar wind. Although the equivalent width of each line is several eV or less,  more than ten lines can contribute an absorption feature observed with CCD resolution. Therefore, the equivalent widths of the absorption features in the spectra of Suzaku J1305$-$4930 are compatible with those observed in Cyg X-1. We also note that some positive residuals remain below 0.9~keV in the spectra,  which vary with time. Interestingly, in two ULXs, NGC~1313 X-1 and NGC~5408 X-1, soft X-ray emission and absorption lines of highly ionized ions, some with significant blue shift, are observed (Pinto et al. 2016). They are interpreted as outflows of the source. Observations with a higher energy resolution is necessary to investigate it, as well as the origin of the absorption features.

Another popular model for this type of hard X-ray transients is a p-free disk model (\verb|diskpbb| in \verb|XSPEC|). We apply it to the spectra instead of the MCD model and find that $p$ is consistent with the nominal value of 0.75, although errors are large. Also, we  examine an additional power-law component with a fixed photon index of 2  and find it  to be  unnecessary.

\renewcommand{\arraystretch}{1}
\begin{table*}[t]
  \caption{Best-fit spectral parameters  with the MCD model.}

 \begin{center}
  \begin{tabular}{cccccccc}
   \hline\hline
  parameters & 2005 1st & 2006 1st & 2010 1st & 2010 2nd & 2010 3rd & 2010 4th & 2011 1st \\ \hline
  $kT_{\rm in}$ (keV) & - & - &	$1.15^{+0.07}_{-0.06}$ &	$1.01\pm0.06$ & 	$0.97\pm0.06$ &	$0.80^{+0.13}_{-0.11}$ & - \\
  $R_{\rm in}$ (km) & - & - &	$64^{+14}_{-11}$ &	$80^{+20}_{-16}$ &	$83^{+22}_{-17}$ &	$69^{+55}_{-30}$ & - \\
  constant & - & - & $0.8\pm0.5$ &	$1.7\pm0.5$ &	$1.4\pm0.4$ &	$1.0\pm0.3$ & - \\
  Flux (10$^{-13}$~erg s$^{-1}$~cm$^{-2}$) & $<$~0.9 & $<$~2.1 &	$5.4\pm0.1$ &	$4.8\pm0.1$ &	$4.4\pm0.1$ &	$1.3\pm0.1$ & $<$~0.4 \\
  $L_{\rm x}$ (10$^{38}$~erg s$^{-1})$ & $<~1.8$ & $<$~4.8 & $8.9^{+0.5}_{-0.5}$ &	$7.9^{+0.4}_{-0.7}$ &	$7.3^{+0.5}_{-0.7}$ &	$2.2^{+0.3}_{-0.8}$ & $<$~2.4 \\
  $L_{\rm disk}$ (10$^{38}$~erg~s$^{-1}$) &  - & - & 9.4 & 8.4 & 7.7 & 2.4 & - \\  \hline
  ${\chi}^2$ / d.o.f. & - & - & 147.6 / 148 &	135.1 / 129 &	148.0 / 128 &	95.0 / 58 & - \\
  \hline\hline
  \end{tabular}

\end{center}
\label{tab:parameter_dbb}
\small{}
\end{table*}


\renewcommand{\arraystretch}{1}
\begin{table*}[t]
\caption{Best-fit parameters with the \texttt{gabs*diskbb} model. }
 \begin{center}

  \begin{tabular}{ccccc}
   \hline\hline
  parameters & 2010 1st & 2010 2nd & 2010 3rd  & 2010 4th \\ \hline
  constant & $1.0\pm0.5$ &	$1.9\pm0.5$ &	$1.6\pm0.4$ &	$1.2\pm0.3$ \\
  Line~$E$ (keV) &	$0.96\pm0.02$ &	$0.94^{+0.04}_{-0.03}$ &	$0.96^{+0.03}_{-0.02}$ &	$0.98\pm0.03$ \\
  Sigma (keV) & $0.005^{+0.035}_{-0.002}$ &	$0.007^{+0.056}_{-0.003}$ &	$0.009^{+0.372}_{-0.004}$ &	$0.018^{+0.052}_{-0.006}$ \\
  Equivalent Width (eV) & 9$^{+98}_{-6}$ & 11$^{+84}_{-5}$ & 39$^{+1608}_{-17}$ & 95$^{+276}_{-32}$ \\
  $kT_{\rm in}$ (keV) &	$1.12^{+0.07}_{-0.06}$ &	$0.9\pm0.1$ &	$0.93\pm0.06$ &	$0.6^{+0.1}_{-0.2}$ \\
  $R_{\rm in}$ (km) &	$68^{+15}_{-13}$ &	$82^{+21}_{-17}$ &	$89^{+26}_{-20}$ &	$116^{+168}_{-62}$ \\ \hline
${\chi}^2$ / d.o.f. &	130.8 / 145 &	130.1 / 127 &	128.0 / 125 &	71.2 / 55 \\
\hline\hline
  \end{tabular}
\end{center}

\label{tab:parameter_gabsdbb}
\end{table*}



\subsection{On-Off Difference Spectra of Suzaku~J1305$-$4930}
\begin{figure}[h]
\begin{center}
\FigureFile(100mm,100mm){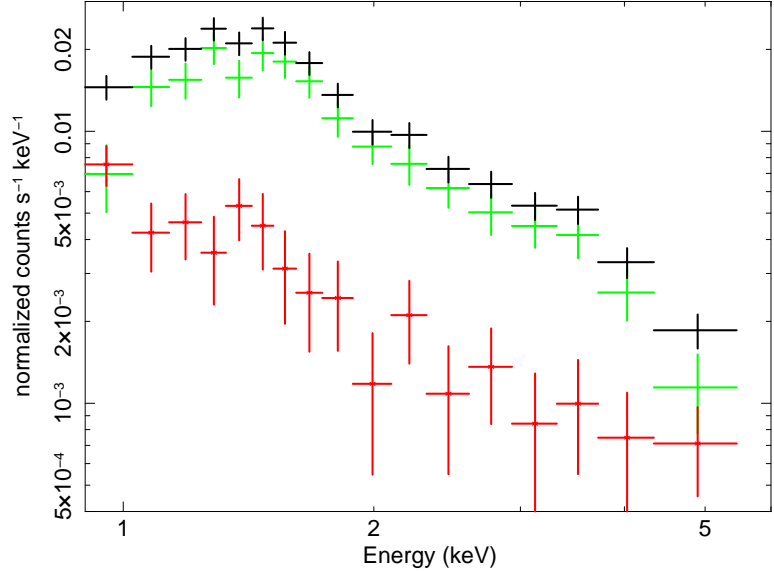}
\end{center}
\caption{Difference spectrum (green), generated by subtracting the 2011 1st spectrum (red) from the 2010 1st spectrum (black).}
\label{fig:subtract}
\end{figure}

We  perform spectral analyses with  the difference spectra of the SOURCE region; the 2011 1st data (off-source), where the transient source is not visible, are subtracted from  the on-source data of the 2010 1st, 2010 2nd, 2010 3rd,  and 2010 4th.  Providing that there is no variability  in any other X-ray sources in the field of view, including the AGN,  these difference spectra should be  those of Suzaku~J1305$-$4930 only. Note that the NXB spectrum  in each observation  is subtracted using \verb|mathpha| in \verb|heasoft|. Figure \ref{fig:subtract} shows the difference spectrum of the 2010 1st  minus 2011 1st, as an example, along with the on-source and off-source spectra. The spectra were binned into energy intervals with  40 events per bin at minimum. \newline \indent Table \ref{tab:parameter_subtract} shows the best-fit parameters  from fitting of the difference spectra with the MCD model. The obtained parameters are consistent with those in table \ref{tab:parameter_gabsdbb} in section \ref{subsec:source}, although errors are larger  with the difference spectra.
This supports validity of our analysis  described in section \ref{subsec:source}.  Hence, we will primarily focus on our results obtained in section \ref{subsec:source} in the following discussion.


\renewcommand{\arraystretch}{1}
\begin{table*}[t]
  \caption{Best-fit parameters  with the MCD model  applied to the difference spectra.}
 \begin{center}
  \begin{tabular}{ccccc}
   \hline\hline
  parameters & 2010 1st & 2010 2nd & 2010 3rd  & 2010 4th \\ \hline
  $kT_{\rm in}$ (keV) &   $1.1^{+0.1}_{-0.1}$ &      $1.0^{+0.1}_{-0.1}$ &  $1.0^{+0.1}_{-0.1}$ &     $0.7^{+0.2}_{-0.2}$ \\
$R_{\rm in}$ (km) &   $71^{+34}_{-23}$ &        $78^{+44}_{-30}$ &        $86^{+46}_{-31}$ &  $84^{+188}_{-75}$ \\
$L_{\rm x}$ (10$^{38}$~erg~s$^{-1}$) & $8.5^{+0.6}_{-1.7}$ &       $8.2^{+0.7}_{-2.2}$ &       $7.5^{+0.5}_{-1.3}$ & $2.2^{+0.2}_{-2.2}$ \\ \hline
${\chi}^2$ / d.o.f. &  8.8 / 15 &       7.1 / 12 & 8.8 / 12 &   14.9 / 25 \\
\hline\hline
  \end{tabular}
\end{center}
\label{tab:parameter_subtract}
\end{table*}

\subsection{X-ray Lightcurve}\label{subsec:light curve}
\begin{figure}[h]
\begin{center}
\FigureFile(100mm,100mm){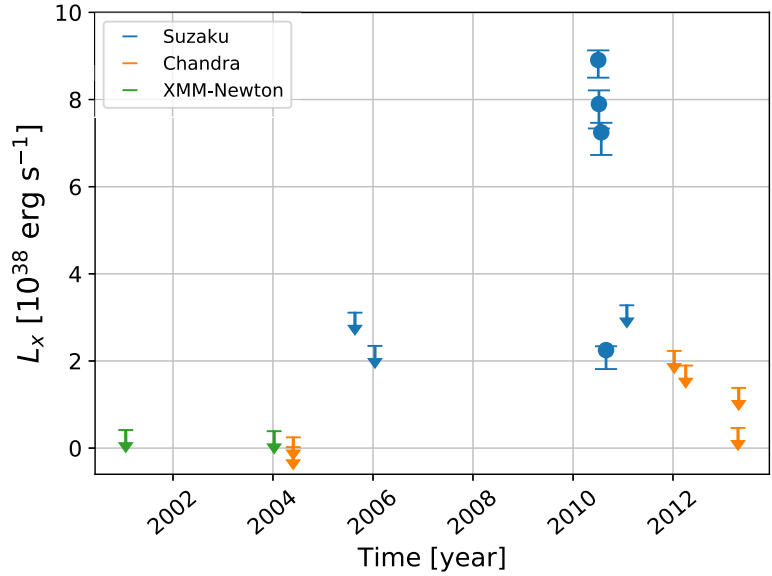}
\end{center}
\vspace{4cm}
\caption{Long-term light curve of the transient source Suzaku~J1305$-$4930 in the SOURCE region. The observations  with Chandra and XMM-Newton provide  upper limits of the luminosity only.  \label{fig:L_all}}
\end{figure}

We apply the same analysis procedure in section \ref{subsec:source}, employing the MCD model, to the data of the 2005 1st, 2006 1st, and 2011 1st, and obtain the upper limits of  ${L_{\rm x}}$. We also analyze the Chandra data of OBSIDs 846, 4899, 4900, 13791, 14412, 14984, and 14985, and those of XMM-Newton of OBSIDs 0112310301 and 0204870101, to derive the ${L_{\rm x}}$ upper limits. The results are summarized in figure \ref{fig:L_all}, showing the long-term light curve of Suzaku~J1305$-$4930 in the 0.3--10~keV band.

\section{Discussion}\label{s5}

\subsection{Estimation of the Black Hole Mass}
\begin{figure*}[h]
\begin{center}
\FigureFile(100mm,100mm){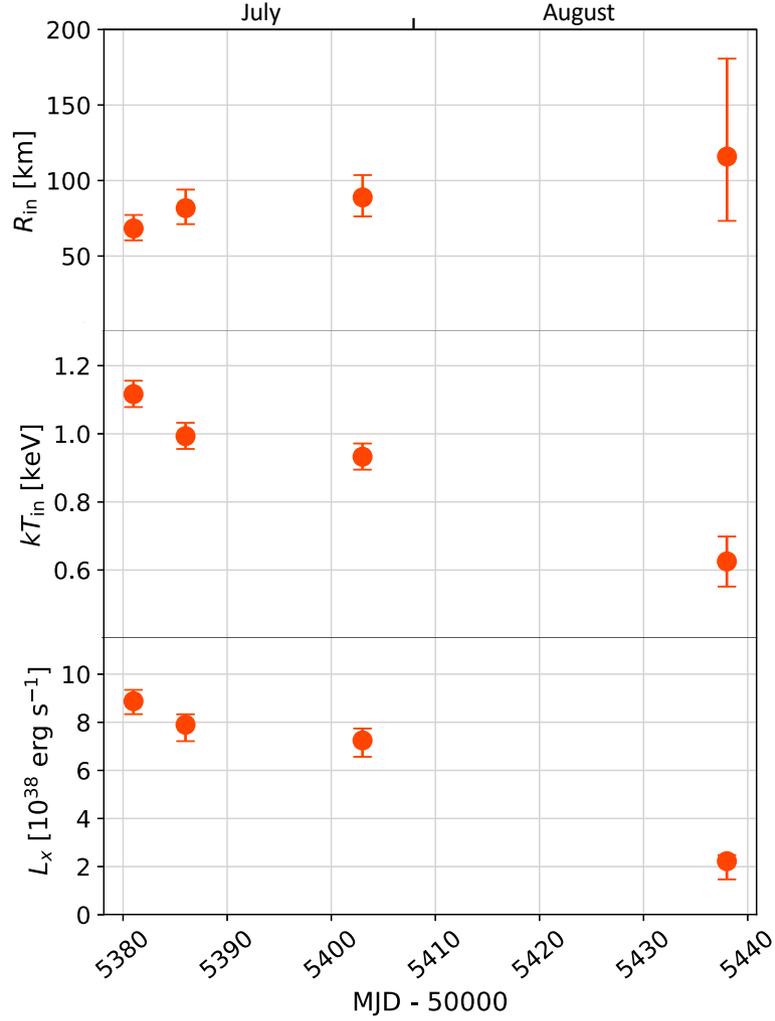}
\end{center}
\vspace{1.5cm}
\caption{Top panel: $R_{\rm in}$ with the assumed inclination angle  of $60^{\circ}$. Middle: $kT_{\rm in}$. Bottom: $L_{\rm x}$ (0.3--10 keV)  for the assumed distance  of 3.72~Mpc.}
\label{fig:parameter_change}
\end{figure*}

The highest observed X-ray luminosity $L_{\rm x}$ of Suzaku~J1305$-$4930 is $(8.9^{+0.2}_{-0.4}) \times 10^{38}$~erg s$^{-1}$ (0.3--10 keV)  for the assumed distance of 3.72~Mpc.  Providing that this  is the Eddington luminosity, the  mass of Suzaku~J1305$-$4930  is estimated to be 7~$M_{\odot}$.
Alternatively,  the innermost disk radius of  $R_{\rm in} = 68^{+9}_{-8} \alpha^{1/2}$ km provides an estimate of the mass of $\sim$8~$\times (\frac{\alpha}{1.0})^{\frac{1}{2}} M_{\odot}$,  providing that $R_{\rm in}$ is equal to the last stable orbit of  the standard accretion disk around a Schwarzschild black hole, i.e., $R_{\rm in} = 3R_{\rm s}$, where $R_{\rm s}$ is the Schwarzschild radius.
 The value of the mass estimated in either way  suggests that this source is a stellar-mass black hole binary.
Figure \ref{fig:parameter_change} shows that $kT_{\rm in}$ decreased from 1.12$\pm0.04$~keV at the 2010 1st to 0.63$\pm0.07$~keV at the 2010 4th in about 60~days. In the same period, its ${L_{\rm x}}$ decreased along with $T_{\rm in}$, while $R_{\rm in}$  did not significantly change at $\sim$75~km. These properties are very similar to those observed in Galactic black hole binaries (e.g. Done et al. 2007).

In reality, the  inner-edge of the disk is not identical to R$_{\rm in}$ of the MCD model and depends on the black-hole spin parameter and inner-boundary conditions.
However, the data statistics of our data are too limited, and various parameters degenerate each other to be constrained from spectral fitting. We thus employ the model of the general-relativistic accretion disk around a Kerr black hole (\texttt{kerrbb} in \texttt{XSPEC}; Li et al. 2005), with fixing the spin parameter at 0, the disk inclination at 60$^{\circ}$, the distance at 3.72~Mpc, and other detailed accretion disk parameters as their default values (e.g. the spectral hardening factor ($T_{\rm col}$/$T_{\rm eff}$) = 1.7) to fit the spectra. We obtain the black hole mass of 6.4--7.9~M$_{\odot}$ with this model. This value is consistent with that evaluated with diskbb model.


\subsection{$L_{\rm disk}$ vs $T_{\rm in}$ Diagram}

\begin{figure*}[h]
\begin{center}
\FigureFile(100mm,100mm){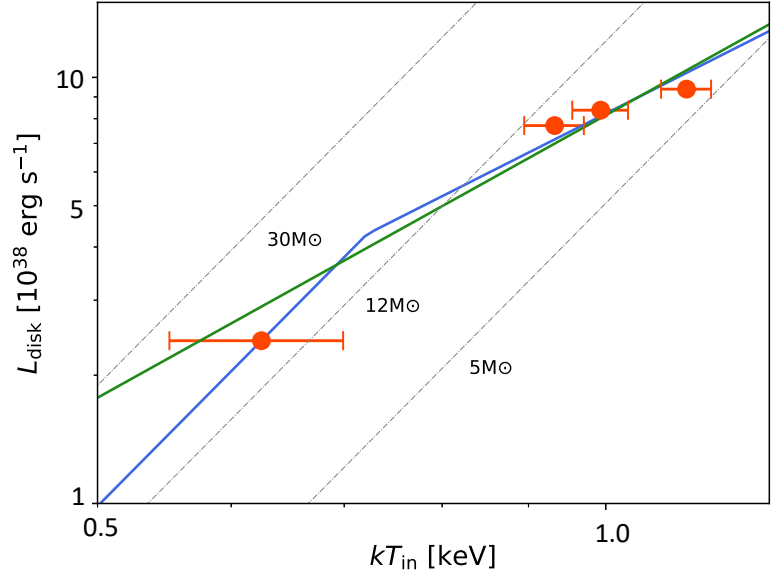}
\end{center}
\vspace{4cm}
\caption{Relation between  $L_{\rm disk}$ and  $kT_{\rm in}$. The  gray lines show $L_{\rm disk} \propto T_{\rm in}^4$ for  some example black-hole masses. Green solid line shows a single power-law model with an index of 2.2, and blue solid line shows a broken power-law model with  (fixed) indices of 2 and 4.}
\label{fig:L_Tin}
\end{figure*}

If Suzaku~J1305$-$4930 is a stellar-mass black-hole binary in NGC~4945, it means that we have serendipitously obtained  multiple observations with relatively short intervals in between of  a source of this class outside our Galaxy.
Figure \ref{fig:L_Tin} shows the $L_{\rm disk}$ vs $T_{\rm in}$ diagram of Suzaku~J1305$-$4930 for the four observations.  The relation of $L_{\rm disk} \propto T_{\rm in}^4$ is expected for  the standard accretion disk (Shakura \& Sunyaev 1973) with a  time-invariable  inner radius of the disk.
However, the $L_{\rm disk}$ vs $T_{\rm in}$ relation is fitted with a single power-law model with the best-fit index of $2.2 \pm 0.5$, which does not support the expected value of 4,  with $\chi^2$ / d.o.f. = 3.7 / 2. The observed index is rather consistent with 2, which is expected for a slim-disk model. Indeed, Watarai et al. (2000) simulated many slim-disk spectra, fitted them with a MCD model, and derived  the empirical relation $R_{\rm in} \propto T_{\rm in}^{-1}$ ($L_{\rm disk} \propto T_{\rm in}^2$). We thus interpret that Suzaku~J1305$-$4930 was  not in  the standard-disk state but in the slim-disk state during the four observations in 2010.

We apply  to the $L_{\rm disk}$ vs $T_{\rm in}$ relation of Suzaku~J1305$-$4930 another model: a broken power-law model with  fixed  indices of 4 and 2 at the lower  and  higher temperature sides, respectively.
  Figure \ref{fig:L_Tin}  shows the best-fit result, which gives $\chi^2$ / d.o.f. = 2.7 / 2. The break point is  determined to be $kT_{\rm in}$ = 0.7$\pm 0.2$~keV, which can be translated  into  a luminosity of (4.3$\pm2.4) \times 10^{38}$~erg s$^{-1}$. So, the 2010 1st, 2010 2nd and 2010 3rd is in slim-disk state, and the 2010 4th is in standard-disk state.
\newline \indent Kubota et al. (2004) and Kubota et al. (2001) reported  state transitions in the Galactic black hole binaries XTE~J1550$-$564 and GRO~J1655$-$40, respectively. Their $L_{\rm disk}$ vs $T_{\rm in}$ diagrams exhibit  clear state transitions from the standard-disk state ($L_{\rm disk} \propto T_{\rm in}^4$) to the slim-disk state ($L_{\rm disk} \propto T_{\rm in}^2)$. Our results on Suzaku~J1305$-$4930 may alternatively be interpreted with a similar state transition.

\section{Summary}

We discover an X-ray transient, named Suzaku~J1305$-$4930, about 3~kpc  away from the nucleus of NGC~4945. Among the seven Suzaku observations of this region, the source was detected in four observations from 2010 July to 2010 August. The highest observed 0.3--10 keV X-ray luminosity $L_{\rm x}$ is $(8.9^{+0.2}_{-0.4})\times 10^{38}$~erg~s$^{-1}$.
The X-ray spectra are well reproduced with a MCD model. The parameters $L_{\rm x}$ and  $R_{\rm in}$ consistently suggest that this source is a stellar-mass black-hole binary with a mass of $\sim$10 $M_{\odot}$. The $L_{\rm disk}$ vs $T_{\rm in}$ relation is, however, not reproduced with the standard $L_{\rm disk} \propto T_{\rm in}^4$ relation; a flatter relation ($L_{\rm disk} \propto T_{\rm in}^{2.2 \pm 0.5}$) or power-law relation with a break is suggested.

\begin{ack}
This work is supported by Japan Society for the Promotion of Science (JSPS) KAKENHI Grant Numbers JP 16H00949, 16K13787, 18K18767, 18J20523, 19K21884, 19H00696, 19H01908, 23340071, 26109506 and also by the Mitsubishi Foundation Research Grants in the Natural Sciences 201910033.
\end{ack}

 \end{document}